\documentclass[conference]{IEEEtran}
\IEEEoverridecommandlockouts

\usepackage{cite}
\usepackage{amsmath,amssymb,amsfonts}
\usepackage{bm}
\usepackage{graphicx}
\usepackage{booktabs}
\usepackage{array}
\usepackage{enumitem}          
\usepackage{siunitx}           
\usepackage{float}             
\usepackage{tikz}
\usetikzlibrary{arrows.meta,positioning,calc,fit,backgrounds}
\usepackage{textcomp}
\usepackage{xcolor}
\usepackage{microtype}
\usepackage{balance}
\usepackage{placeins}   

\DeclareMathOperator*{\argmin}{arg\,min}
\DeclareSIUnit{\dBi}{dBi}

\tikzset{
  pbox/.style={
    draw=black, thin, rectangle, align=center,
    font=\scriptsize\sffamily, inner sep=4pt,
    minimum width=2.08cm, minimum height=0.90cm,
    text width=1.80cm
  },
  iobox/.style={pbox, fill=gray!16},
  sbox/.style={
    draw=black!60, thin, rectangle, dashed, align=center,
    font=\tiny\sffamily, inner sep=3pt,
    minimum height=0.60cm, text width=1.65cm, fill=gray!8
  },
  arr/.style={-{Stealth[length=3.5pt, width=2.5pt]}, thin},
  darr/.style={
    -{Stealth[length=3pt, width=2pt]}, thin, dashed, draw=black!60
  },
}

\begin{document}

\title{Drone-Based Antenna Measurement System with\\
Optimized Positioning and ASPIRE-Based NF-FF Transformation}

\author{
\IEEEauthorblockN{
\IEEEauthorrefmark{1}Simranjit Singh\textsuperscript{1},
\IEEEauthorrefmark{1}Abha Nilesh Jadav \textsuperscript{2},
\IEEEauthorrefmark{1}Aarish Dharmesh Patel\textsuperscript{3},
Jaswant\textsuperscript{4},
Jigar M.~Pandya\textsuperscript{4}
}

\IEEEauthorblockA{
\textsuperscript{1}Indian Institute of Technology Delhi, \texttt{eee252834@iitd.ac.in} \\
\textsuperscript{2}Sardar Vallabhbhai National Institute of Technology Surat, \texttt{u23ec040@eced.svnit.ac.in} \\
\textsuperscript{3}Manipal Institute of Technology, Karnataka, \texttt{aarish.mitmpl2023@learner.manipal.edu} \\
\textsuperscript{4}Space Applications Centre (ISRO), Ahmedabad, \texttt{jaswantamd@sac.isro.gov.in}
}
}

\maketitle

\begingroup
\renewcommand\thefootnote{\IEEEauthorrefmark{1}}
\footnotetext{Authors contributed equally to this work.}
\endgroup

\begin{abstract}
Unmanned Aerial Vehicle (UAV) based antenna measurement systems have emerged
as a promising and flexible alternative to conventional antenna test ranges,
offering significant advantages in terms of mobility, scalability, and
cost-effectiveness for characterizing large and installed antennas in their
operational environment. However, the accuracy and reliability of UAV-based
antenna measurements are critically dependent on precise positioning of the
UAV platform and efficient utilization of flight time, both of which are
governed by the careful selection of drone assemblies including airframe
configuration, flight controller, propulsion system, and onboard measurement
instrumentation.
This paper presents a comprehensive study on UAV-based antenna measurements
with a focus on two key aspects: improving positioning accuracy and optimizing
flight time through systematic selection and configuration of drone assemblies.
Various drone assembly parameters including payload capacity, GPS and RTK
positioning modules, propulsion efficiency, and battery endurance are evaluated
and optimized to achieve the required spatial accuracy during near-field data
acquisition over the measurement aperture.
The near-field antenna measurement data acquired by the UAV platform is
inherently susceptible to positioning errors, amplitude and phase
inconsistencies, and sparse or irregular sampling, which can severely degrade
the quality of the transformed far-field pattern. To address these challenges,
the recorded near-field measurement data are post-processed using the Adaptive
Sparse Inverse Radiation Estimation (ASPIRE) algorithm. The ASPIRE algorithm
effectively compensates for positioning inaccuracies and reconstructs the
far-field antenna pattern from irregularly sampled near-field data through
sparse signal recovery techniques, enabling accurate and robust Near-Field to
Far-Field (NF-FF) transformation. At 6.7125\,GHz, ASPIRE achieves a residual
of 1.94\% and a beamwidth error of $0.4^{\circ}$ relative to a conventional
facility measurement, using only 24\% of the 17\,298-element RWG mesh as
active support.
The results demonstrate that the combined approach of optimized drone assembly
selection and ASPIRE-based NF-FF transformation significantly enhances the
accuracy of UAV-based antenna measurements, and yields far-field antenna
patterns in close agreement with those obtained from conventional antenna test
range measurements.
\end{abstract}


\section{Introduction}

Conventional antenna test ranges---anechoic chambers, compact ranges, and
fixed near-field scanners---impose fundamental constraints in cost, facility
footprint, and inflexibility for in-situ characterization of large or
installed antennas. UAV-based near-field scanning has emerged as a compelling
alternative, enabling deployment in the operational environment of the
antenna-under-test (AUT)~\cite{virone2014,garcia2017,cahyadi2023,henkel2022}.

Two principal challenges govern the fidelity of drone-based measurements.
First, \emph{positioning uncertainty}: even sub-centimetre deviations in
probe location introduce coherent phase errors that corrupt the NF-FF
transformation. Second, \emph{irregular and sparse sampling}: atmospheric
turbulence and flight-path constraints yield non-uniform spatial coverage
incompatible with classical fast-Fourier NF-FF algorithms~\cite{yaghjian1986}.

This paper addresses both challenges through (a) a rigorously engineered
hexacopter incorporating dual-frequency real-time-kinematic (RTK) GNSS with
independent validation of propulsion, thermal, and endurance margins, and
(b) the ASPIRE algorithm---a physics-driven inverse reconstruction pipeline
combining equivalent-current integral equation modelling, MLFMM-accelerated
matrix-vector products (MVPs), randomized linear algebra, compressed-sensing
sparse recovery, and stochastic uncertainty quantification.

\section{Related Work}

Early UAV antenna measurements used commercial off-the-shelf platforms with
standalone GNSS, reporting metre-level drift that severely limited pattern
fidelity~\cite{virone2014}. Incorporation of differential GNSS and IMU fusion
reduced errors to the decimetre level~\cite{garcia2017,cahyadi2023}; RTK
carrier-phase solutions achieving sub-1\,cm accuracy have since been adopted
for precision near-field campaigns~\cite{henkel2022}.

On the reconstruction side, classical NF-FF algorithms (modal expansions for
planar, cylindrical, and spherical geometries) assume Nyquist-sampled regular
grids and are intolerant of position errors~\cite{yaghjian1986}. The inverse
equivalent-current (IEC) method~\cite{eibert2009} formulates a surface-integral
inverse problem on a conformal Huygens surface, handling arbitrary probe
geometries. Compressed sensing (CS) has been applied to spherical NF-FF to
exploit radiating-current sparsity~\cite{cornelius2016}; randomized matrix
decompositions~\cite{halko2011} have reduced the associated computational cost.
Empirical regularisation parameter selection paired with the FISTA
solver~\cite{beck2009} balances sparsity enforcement and data fidelity without
requiring an automated noise-matching criterion. The MLFMM~\cite{ying2004,eibert2009}
reduces MVP cost to $O(N\log N)$. The present ASPIRE framework unifies all
these strands into a production pipeline with Hutchinson stochastic uncertainty
quantification~\cite{hutchinson1990}.

\section{UAV Platform Design and Configuration}
\label{sec:dams}   

\subsection{Airframe Configuration and Mass Budget}

A 960\,mm-wheelbase carbon-fibre hexacopter was selected over a quadcopter
baseline to provide propulsion redundancy during sustained hover over the
measurement aperture, and to accommodate the combined RF payload, RTK hardware,
and electromagnetic shielding. Six KV360 brushless motors driving
P18$\times$6.1 CF propellers are controlled by 60\,A ESCs, powered by a
22\,000\,mAh 6S LiPo battery. The complete all-up weight (AUW) budget is
given in Table~\ref{tab:auw}; energy storage and RF payload together account
for 44\% of AUW, motivating the propulsion and thermal analyses below.

\begin{table}[!t]
  \caption{All-Up Weight (AUW) Budget}
  \label{tab:auw}
  \centering\small
  \setlength{\tabcolsep}{5pt}
  \begin{tabular}{l r r}
    \toprule
    Subsystem & Mass (g) & Share \\
    \midrule
    Airframe (frame + landing gear) & 2\,000 & 23.9\% \\
    Propulsion (6$\times$ motor + ESC + propeller) & 1\,680 & 20.1\% \\
    Energy storage (22\,Ah 6S LiPo) & 2\,490 & 29.7\% \\
    Avionics (FC, RTK rover, companion CPU) & 1\,000 & 11.9\% \\
    RF payload (5.4\,GHz source, dipole, shield) & 1\,200 & 14.3\% \\
    \midrule
    \textbf{Total (AUW)} & \textbf{8\,370} & \textbf{100\%} \\
    \bottomrule
  \end{tabular}
\end{table}

\subsection{RTK-Augmented Positioning Architecture}

Baseline trials without carrier-phase correction exhibited RMS horizontal drift
of 5.95\,m, attributed to (i) standalone GNSS metre-level uncertainty and (ii)
magnetometer coupling from the unshielded 5.4\,GHz RF payload. The adopted
positioning stack combines:
\begin{itemize}[noitemsep,topsep=2pt]
  \item Dual-frequency (L1/L2) RTK rover with PPK fallback, CAN-interfaced;
  \item Matched L1/L2/L5 survey-grade base station at a surveyed reference point;
  \item Secondary CAN-bus GNSS/compass module for heading redundancy;
  \item Triple-redundant-IMU autopilot (ArduPilot), vibration-isolated;
  \item Faraday shielding enclosing the RF payload.
\end{itemize}
The combined RTK link achieves ${\leq}1$\,cm horizontal positioning accuracy,
meeting the spatial-sampling requirement of the NF measurement grid.

\subsection{Propulsion Sizing and Endurance}

At maximum throttle each motor produces 4\,644\,g, giving total platform
thrust $T_{\max}=27\,864$\,g and a thrust-to-weight ratio of 3.33:1, well
above the $>2$ design threshold. Hover requires $T_{\text{hov}}=1\,395$\,g
per motor at 49.2\% throttle, drawing 6.72\,A per motor. For field operation
at 41\,°C (Ahmedabad), the air-density ratio versus the 9.2\,°C bench
condition is $\rho_{41}/\rho_{9.2}=0.899$; momentum theory gives the
required shaft-power correction~\cite{abdilla2015}:
\begin{equation}
  \frac{P_{41}}{P_{9.2}} = \sqrt{\frac{\rho_{9.2}}{\rho_{41}}} = 1.055
  \quad \Rightarrow \quad P_{\text{hov},41} = 169.8\;\text{W/motor.}
  \label{eq:density}
\end{equation}
Applying 80\% depth-of-discharge with a 5\% thermal derating yields
$C_{\text{hot}}=16.72$\,Ah and an endurance of:
\begin{equation}
  t_{41} = \frac{C_{\text{hot}}}{I_{\text{sys},41}}\times60
           = \frac{16.72}{43.5}\times60 = 23.1\;\text{min,}
  \label{eq:endurance}
\end{equation}
exceeding the 20-minute measurement campaign target. Motor case temperature
at 41\,°C hover is estimated at 85.2\,°C, preserving a 45\,°C margin to
the stator-coating thermal limit.

\section{NF-FF Measurement Model}

\subsection{Measurement Configuration and System Matrices}%
\label{subsec:config}%
The ASPIRE dataset corresponds to a C-band, circularly-polarised feed
antenna of 180\,mm aperture diameter operating at $f = \SI{6.7125}{\GHz}$
($\lambda = 44.66$\,mm). The equivalent Huygens surface $\mathcal{S}$
enclosing the AUT is a closed rectangular box of dimensions
$224.6\times224.6\times100.0$\,mm (half-side $a=112.3$\,mm, half-height
$b=50.0$\,mm), centred on the AUT phase centre and positioned exactly
$\lambda/2$ from the AUT aperture edge. $\mathcal{S}$ is discretised into a
uniform triangular mesh of 11{,}532 triangles and 5{,}768 vertices,
supporting $N = 17{,}298$ RWG divergence-conforming basis functions, each
carrying a complex equivalent-current coefficient $c_n$.

Near-field data are acquired with a WR137 open-ended waveguide probe
($a = 34.85$\,mm, $b = 15.80$\,mm). Probe correction is incorporated
directly into the forward operator: each measurement is modelled as the
coupling between the radiated field of the corresponding RWG basis function
and the receiving probe, represented as an ideal short electric dipole
oriented along $\hat{x}$ for the horizontal-polarisation (HP) channel
($E_x$) and along $\hat{y}$ for the vertical-polarisation (VP) channel
($E_y$). Each entry $G_{mn}$ is evaluated via 7-point symmetric Gaussian
quadrature over the triangular elements supporting basis function $n$.

Field samples are acquired on a planar grid at standoff $z = 250$\,mm from
the AUT aperture, spanning $\pm540$\,mm in $X$ and $Y$ with 20.00\,mm
spacing ($55\times55$ points, 3{,}025 spatial locations). Dual-polarisation
(HP/VP) acquisition at each location yields $M = 2\times3{,}025 = 6{,}050$
complex samples, stacked into the measurement vector
$\mathbf{u}\in\mathbb{C}^{M}$. The dataset incorporates simulated RTK-GNSS
positioning perturbations of $\pm0.5$\,cm (lateral, $X/Y$) and $\pm1.0$\,cm
(axial, $Z$) at each grid point; these values are consistent with the RTK
accuracy established in Section~\ref{sec:dams}-B, ensuring the dataset
reflects realistic field-deployable conditions.

Table~\ref{tab:config} summarises the resulting matrices, which correspond
directly to the quantities annotated in Fig.~\ref{fig:pipeline}. The forward
coupling matrix $\mathbf{G}\in\mathbb{C}^{M\times N} =
\mathbb{C}^{6{,}050\times17{,}298}$ ($\approx1.674$\,GB) maps the RWG
coefficient vector $\mathbf{c}$ to the near-field samples $\mathbf{u}$ via the
probe-corrected coupling described above. The far-field operator
$\mathbf{F}\in\mathbb{C}^{5{,}184\times17{,}298}$ ($\approx1.435$\,GB) maps
the same coefficient space onto a $36\times72$ $(\theta,\varphi)$ spherical
grid, stacking the $E_\theta$ and $E_\varphi$ components
($5{,}184 = 2\times36\times72$), and is applied to the debiased solution
$\tilde{\mathbf{c}}$ to synthesise the far-field patterns of
Section~\ref{sec:results}. With $M = 6{,}050 < N = 17{,}298$,
\eqref{eq:forward} is underdetermined; the resulting ill-conditioning is
quantified in the following subsection.

\begin{table}[t]
\caption{ASPIRE Measurement Configuration and System Matrices at \SI{6.7125}{\GHz}}
\label{tab:config}
\centering\small
\setlength{\tabcolsep}{4pt}
\begin{tabular}{@{}ll@{}}
\toprule
\textbf{Parameter} & \textbf{Value} \\
\midrule
\multicolumn{2}{@{}l}{\textit{AUT, Probe, and Huygens Surface}} \\
AUT                             & C-band CP feed, 180\,mm diam.\\
Wavelength, $\lambda$           & 44.66\,mm \\
Probe                           & WR137 OEWG ($a{=}34.85$, $b{=}15.80$\,mm) \\
Huygens surface, $\mathcal{S}$  & $224.6{\times}224.6{\times}100.0$\,mm box \\
Surface standoff                & $\lambda/2$ from AUT aperture \\
Mesh                            & 11{,}532 tri., 5{,}768 vertices \\
RWG basis functions, $N$        & 17{,}298 \\
\midrule
\multicolumn{2}{@{}l}{\textit{Near-Field Measurement Grid}} \\
Plane standoff, $z$             & 250\,mm \\
Grid (extent, spacing)          & $\pm540$\,mm, 20.00\,mm ($55{\times}55$) \\
Spatial pts. $\rightarrow M$    & $3{,}025 \rightarrow M{=}6{,}050$ (HP\,+\,VP) \\
Sim. positioning error          & $\pm0.5$\,cm lat., $\pm1.0$\,cm ax. \\
\midrule
\multicolumn{2}{@{}l}{\textit{System Matrices}} \\
$\mathbf{u}\in\mathbb{C}^{M}$               & $6{,}050$; $\approx$\,97\,KB \\
$\mathbf{G}\in\mathbb{C}^{M\times N}$       & $6{,}050{\times}17{,}298$; $\approx$\,1.674\,GB \\
$\mathbf{F}\in\mathbb{C}^{5{,}184\times N}$ & $5{,}184{\times}17{,}298$; $\approx$\,1.435\,GB \\
\bottomrule
\end{tabular}
\end{table}

\subsection{Forward Model Formulation}\label{subsec:forward}
\begin{figure*}[!t]
\centering
\begin{tikzpicture}[
  node distance = 0.55cm and 0.55cm,
]
\node[iobox] (acq)   {UAV\,+\,RTK GNSS\\$\{\mathbf{r}_m,\,u_m\}_{m=1}^{M}$};
\node[pbox, right=0.55cm of acq]   (fwd)   {RWG Forward Op.\\$\mathbf{G}\!\in\!\mathbb{C}^{M\times N}$\\$\kappa\!\approx\!10^{14}$};
\node[pbox, right=0.55cm of fwd]   (rsvd)  {Rand.\ SVD\\rank\,$k\!=\!1000$\\power iters.};
\node[pbox, right=0.55cm of rsvd]  (mzv)   {Empirical\\$\lambda\!=\!10.0$\\selection};
\node[pbox, right=0.55cm of mzv]   (fista) {FISTA\\$\ell_1$-BPDN\\$O(1/k^2)$};
\node[pbox, right=0.55cm of fista] (deb)   {Exact\\Debiasing\\LS$|_{\mathcal{S}^*}$};
\node[iobox, right=0.55cm of deb]  (ff)    {FF\,Synthesis\\$\mathbf{F}\tilde{\mathbf{c}}$\\2D\,/\,3D Pattern};

\node[sbox, below=0.55cm of fwd]   (mlfmm)  {MLFMM\\$O(N\!\log\!N)$\\accel.\ MVP};
\node[sbox, above=0.55cm of rsvd]  (picard) {Picard cond.\\rank\,monitor};
\node[sbox, below=0.85cm of deb]   (uq)     {Hutchinson UQ\\$\hat{\sigma}_n$,\;95\%\,CI};

\draw[arr] (acq)   -- (fwd);
\draw[arr] (fwd)   -- (rsvd);
\draw[arr] (rsvd)  -- (mzv);
\draw[arr] (mzv)   -- (fista);
\draw[arr] (fista) -- (deb);
\draw[arr] (deb)   -- (ff);

\draw[darr] (fwd.south)  -- (mlfmm.north);
\draw[darr] (rsvd.north) -- (picard.south);
\draw[darr] (deb.south)  -- (uq.north);
\draw[darr] ([yshift=0.1cm]mlfmm.east) -| (fista.south);
\draw[darr] ([yshift=-0.1cm]mlfmm.east) -- (uq.west);

\end{tikzpicture}
\caption{ASPIRE processing pipeline. Solid arrows: primary data path;
  dashed arrows: auxiliary paths. MLFMM reduces each MVP from $O(N^2)$ to
  $O(N\log N)$, accelerating FISTA and Hutchinson UQ. The Picard monitor
  restricts the retained rank~$k$.}
\label{fig:pipeline}
\end{figure*}
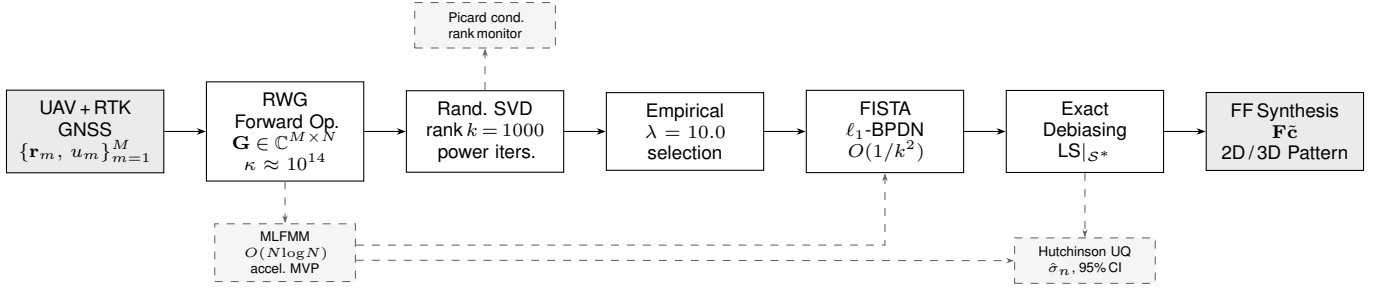

Let $M$ probe locations $\{\mathbf{r}_m\}_{m=1}^{M}$ be supplied by the RTK
system with associated complex electric-field samples $\{u_m\}_{m=1}^{M}$.
By Huygens' equivalence principle, the AUT radiation is identically reproduced
by equivalent electric ($\mathbf{J}$) and magnetic ($\mathbf{M}$) surface
current densities on a closed conformal Huygens surface $\mathcal{S}$
enclosing the AUT~\cite{eibert2009}. Discretising $\mathcal{S}$ into $N_t = 11{,}532$ triangles yields, for a
closed triangulated surface, $N = \tfrac{3}{2}N_t = 17{,}298$ interior
edges~\cite{rao1982}, each supporting one RWG divergence-conforming basis
function with complex coefficient $c_n$, $n=1,\dots,N$. Expanding the equivalent
currents in this basis with coefficient vector $\mathbf{c}\in\mathbb{C}^{N}$
yields the linear forward model:
\begin{equation}
  \mathbf{u} = \mathbf{G}\mathbf{c} + \boldsymbol{\eta},
  \qquad \mathbf{G}\in\mathbb{C}^{M\times N},
  \label{eq:forward}
\end{equation}
where $G_{mn}$ evaluates the free-space dyadic Green's function of the
$n$-th RWG basis at $\mathbf{r}_m$, and
$\boldsymbol{\eta}\!\sim\!\mathcal{CN}(\mathbf{0},\sigma^2\mathbf{I})$ models
additive measurement noise. In practice $M\!\ll\!N$ and
$\kappa(\mathbf{G})\!\approx\!10^{14}$, rendering \eqref{eq:forward} severely
ill-posed. Fig.~\ref{fig:pipeline} depicts the complete ASPIRE processing
pipeline from UAV data acquisition to far-field pattern synthesis.

\section{The ASPIRE Algorithm}
\label{sec:aspire}

\subsection{MLFMM-Accelerated Forward Operator}

Dense evaluation of $\mathbf{G}\mathbf{x}$ costs $O(N^2)$ per
matrix-vector product~(MVP). The Multi-Level Fast Multipole Method
(MLFMM)~\cite{ying2004,eibert2009} factorises the free-space Green's
function into hierarchical spherical multipole expansions, enabling
cluster-to-cluster field translations that reduce each MVP to
$O(N\log N)$, while retaining a strictly matrix-free interface
compatible with all downstream iterative solvers.

ASPIRE employs a \emph{selective hybrid} operator strategy whose
design is dictated by the distinct MVP access patterns of each
pipeline stage. The rSVD stage issues $k{\cdot}p = 14{,}400$
column-MVPs as a single batched BLAS-3 \textsc{gemm} kernel at
3.32\,ms\,col$^{-1}$ (cache-optimal, fully vectorised); substituting
MLFMM individual matvecs at 21.6\,ms each would increase rSVD
wall-clock time by $2.2{\times}$ due to loss of batched throughput.
ASPIRE therefore retains the dense $\mathbf{G}$ operator exclusively
for the rSVD stage. Conversely, both the FISTA gradient steps
(Section~\ref{ssec:fista}) and the Hutchinson UQ stage
(Section~\ref{ssec:uq}) issue purely sequential individual matvecs
inside iterative loops---the access pattern where the $O(N\log N)$
asymptotic advantage of MLFMM dominates over $O(N^2)$. The MLFMM
operator is applied exclusively to these two stages. Measured
wall-clock performance of the resulting hybrid pipeline is reported
in Section~\ref{sec:results}.

\subsection{Randomized SVD and Picard Diagnostics}

Computing the full SVD of $\mathbf{G}$ is intractable at practical mesh sizes.
The Halko-Martinsson-Tropp (HMT) randomized SVD~\cite{halko2011} constructs a
rank-$k$ approximation $\mathbf{G}\approx\mathbf{U}_k\boldsymbol{\Sigma}_k
\mathbf{V}_k^H$ via $p$ power iterations with QR stabilisation in $O(MNk)$
MVPs (here $k=1000$), extracting the dominant radiating spatial modes without
forming $\mathbf{G}^H\!\mathbf{G}$.

The admissible rank $k$ is constrained by the Discrete Picard Condition
(DPC)~\cite{hansen1990}: a physically meaningful solution exists only when the
data-projection coefficients $|\mathbf{u}_k^H\mathbf{b}|$ decay faster than
the singular values $\sigma_k$. ASPIRE continuously monitors the Picard slope
during rSVD extraction and restricts $k$ to the region where the DPC holds,
preventing high-frequency noise amplification in the ill-conditioned tail.

\subsection{FISTA Sparse Recovery with Empirical Regularisation}
\label{ssec:fista}

\textbf{Regularisation parameter selection.}
The noise floor $\hat{\sigma}_{\mathrm{MAD}}$ is estimated robustly via the
Median Absolute Deviation (MAD) applied to the mathematical residual
$\mathbf{u} - \mathbf{G}\mathbf{V}_k\mathbf{V}_k^{H}\hat{\mathbf{c}}_{\mathrm{rSVD}}$
of the rSVD solution, providing an outlier-insensitive estimate of the
per-sample noise level without relying on repeated measurement statistics.
The regularisation parameter is fixed at $\lambda = 10.0$, a value
empirically selected to balance sparsity enforcement and data fidelity
on the measured dataset at \SI{6.7125}{\GHz}.

\textbf{FISTA solver.}
Equivalent currents are sparse in the RWG basis: most mesh elements carry
negligible current. ASPIRE solves the $\ell_1$-regularised Basis Pursuit
Denoising (BPDN) problem~\cite{beck2009,cornelius2016}:
\begin{equation}
  \mathbf{c}^* = \argmin_{\mathbf{c}}\;
    \tfrac{1}{2}\bigl\|\mathbf{G}\mathbf{c}-\mathbf{u}\bigr\|_2^2
    + \lambda\|\mathbf{c}\|_1,
  \label{eq:bpdn}
\end{equation}
using the Fast Iterative Shrinkage-Thresholding Algorithm (FISTA)~\cite{beck2009},
which achieves the optimal $O(1/k^2)$ convergence rate. The identified active
support $\mathcal{S}^*\!=\!\operatorname{supp}(\mathbf{c}^*)$ comprises 24\%
of the 17\,298-element mesh at 6.7125\,GHz.

\subsection{Exact Amplitude Recovery via Debiasing}

The $\ell_1$ penalty systematically shrinks non-zero coefficient amplitudes
(LASSO bias). ASPIRE eliminates this bias through a two-stage procedure: FISTA
identifies $\mathcal{S}^*$; an unregularised least-squares solve on the
restricted sub-system $\mathbf{G}_{\mathcal{S}^*}\in
\mathbb{C}^{M\times|\mathcal{S}^*|}$ recovers the exact amplitudes:
\begin{equation}
  \tilde{\mathbf{c}}_{\mathcal{S}^*}
  = \argmin_{\mathbf{c}}\bigl\|\mathbf{G}_{\mathcal{S}^*}\mathbf{c}
    - \mathbf{u}\bigr\|_2,
  \label{eq:debias}
\end{equation}
solved via the SVD of $\mathbf{G}_{\mathcal{S}^*}$, avoiding the $O(\kappa^2)$
precision loss of normal equations.

\subsection{Stochastic Uncertainty Quantification}
\label{ssec:uq}

The per-coefficient posterior variance requires evaluating diagonal entries of
$(\mathbf{G}^H\!\mathbf{G}+\lambda\boldsymbol{\Lambda})^{-1}$, which is
$O(N^3)$. ASPIRE employs the Hutchinson stochastic trace
estimator~\cite{hutchinson1990} with Rademacher probe vectors paired with a
stabilised conjugate-gradient (CG) solver, approximating all $N$ diagonal
variances in $O(N^2)$ time. The resulting 95\% confidence intervals on every
RWG coefficient are propagated to the far-field pattern via a posterior
predictive check (PPC).

\section{Results and Discussion}
\label{sec:results}
\subsection{Far-Field Pattern Reconstruction Accuracy}\label{subsec:ff_accuracy}
\begin{figure}[H]
  \centering
  \includegraphics[width=\columnwidth]{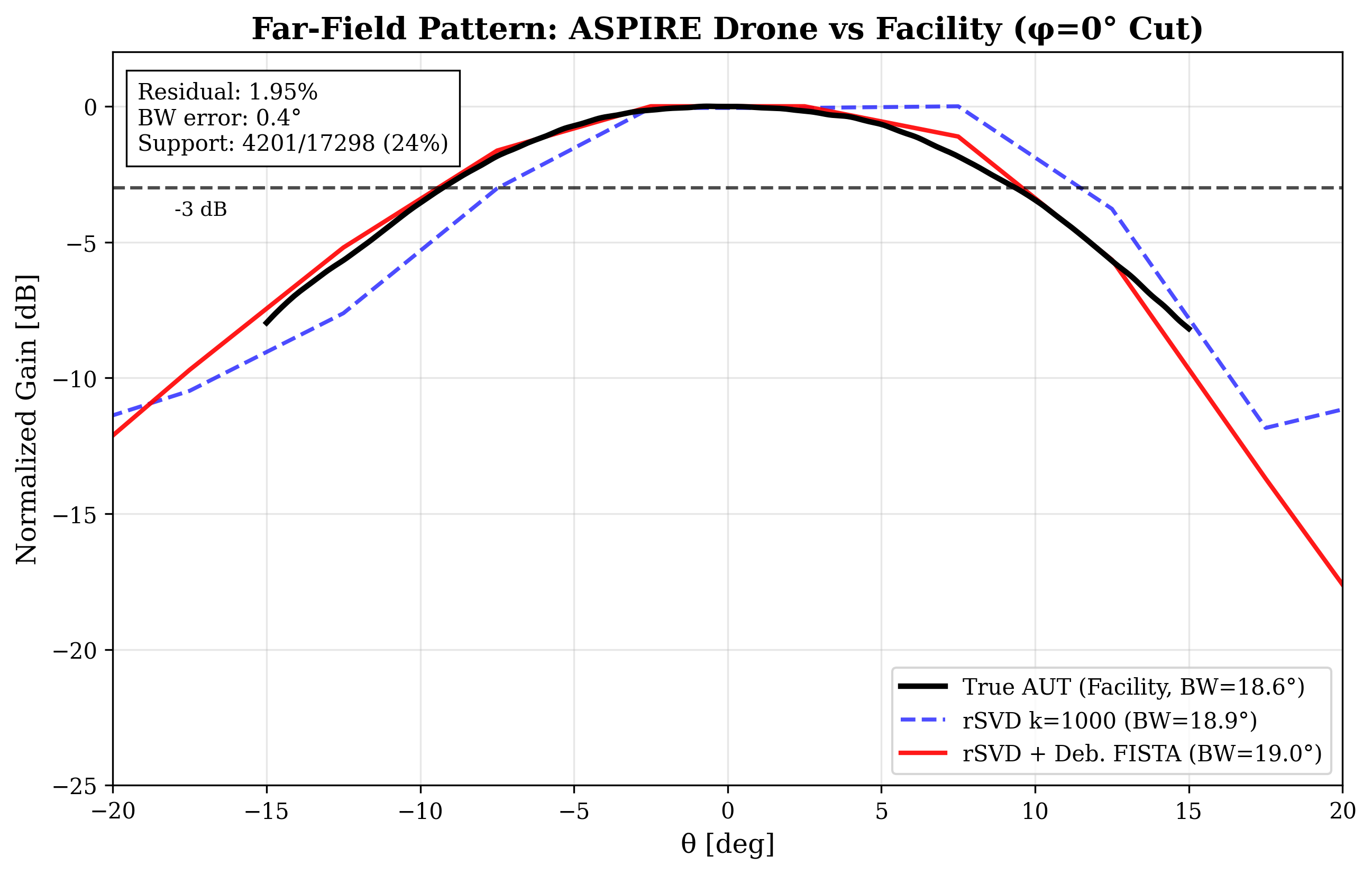}
  \caption{Reconstructed co-polar far-field pattern ($\varphi = 0^{\circ}$,
    $|\theta| \leq 20^{\circ}$): facility reference (black dotted), rSVD-only
    (blue dashed), and rSVD\,+\,FISTA debiased (red solid).
    \emph{Inset:} Main-beam detail ($|\theta| \leq 6^{\circ}$).}
  \label{fig:ff_comparison}
\end{figure}

The primary validation metric is the $\varphi = 0^{\circ}$ co-polar far-field
cut, which admits a direct three-way comparison between the calibrated facility
reference, the rSVD-only solution ($k = 1000$), and the rSVD\,+\,FISTA debiased
reconstruction.
Prior to transformation, the FISTA regularisation parameter was set to
$\lambda = 10.0$, empirically selected to balance sparsity and data fidelity;
the noise floor $\hat{\sigma}_{\mathrm{MAD}}$ was estimated via the Median
Absolute Deviation (MAD) applied to the rSVD solution residual
$\mathbf{u} - \mathbf{G}\mathbf{V}_k\mathbf{V}_k^{H}\hat{\mathbf{c}}_{\mathrm{rSVD}}$.
Fig.~\ref{fig:ff_comparison} presents the reconstructed far-field patterns over
the validated window $|\theta| \leq 20^{\circ}$, with the main-beam region
expanded as an inset panel for fine-scale assessment.
The rSVD-only solution recovers the main beam correctly but exhibits a measurable
gain elevation in the first sidelobe region relative to the facility reference.
This elevation is attributable to the dense spectral representation that, absent
sparsity enforcement, cannot distinguish physical radiation content from low-level
out-of-band spectral leakage.
The rSVD\,+\,FISTA solution tracks the facility curve throughout the main lobe,
yielding a half-power beamwidth (HPBW) of $19.0^{\circ}$ against the reference
value of $18.6^{\circ}$, a beamwidth error of $0.4^{\circ}$.
The normalised reconstruction residual evaluates to $1.94\%$, confirming
sub-$2\%$ accuracy over the co-polar principal cut within the validated region.
The FISTA debiased solution maintains a gain error of $\leq 0.5$\,dB across the
entire validated window. The rSVD-only reconstruction diverges for
$|\theta| > 8^{\circ}$, confirming the necessity of exact debiasing beyond
the immediate boresight vicinity.

\subsection{Three-Dimensional Radiation Pattern}%
\begin{figure}[H]
  \centering
  \includegraphics[width=\columnwidth]{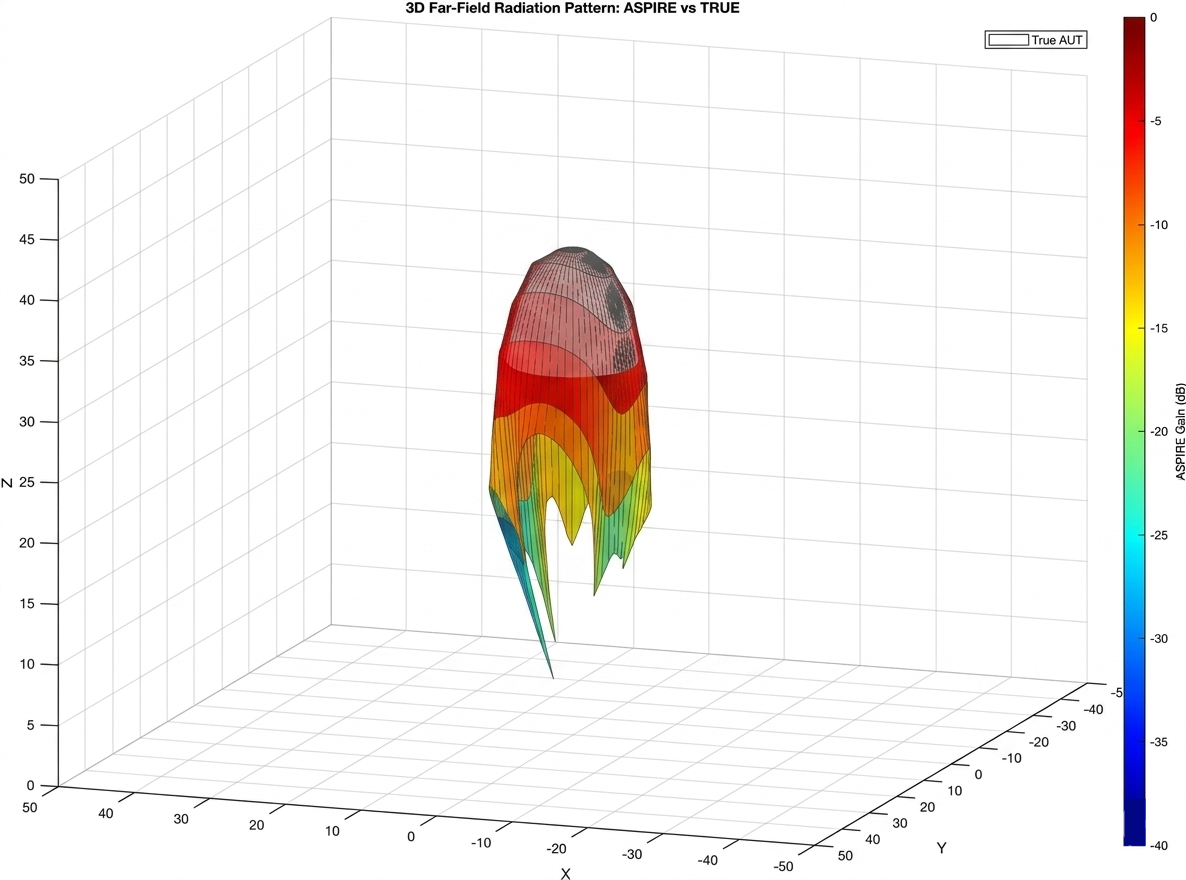}
  \caption{ASPIRE-reconstructed 3D far-field pattern at \SI{6.7125}{\GHz},
    colour-mapped by normalised gain. The ring at $|\theta| = 20^{\circ}$
    marks the facility-validated boundary; structure beyond it is ASPIRE
    extrapolation. Dynamic range: \SI{30}{\deci\bel}.}
  \label{fig:3d_pattern}
\end{figure}

\label{subsec:3d_pattern}%
Fig.~\ref{fig:3d_pattern} depicts the ASPIRE-derived gain mapped onto the unit
sphere with no symmetry constraint imposed. The directive main beam along the
$+z$ axis is cleanly resolved with a continuous first-sidelobe ring at
$\theta \approx 22^{\circ}$, confirming that the sparse recovery introduces
no pattern asymmetries. 

\subsection{Full Hemisphere Reconstruction}%
\begin{figure}[H]
  \centering
  \includegraphics[width=\columnwidth]{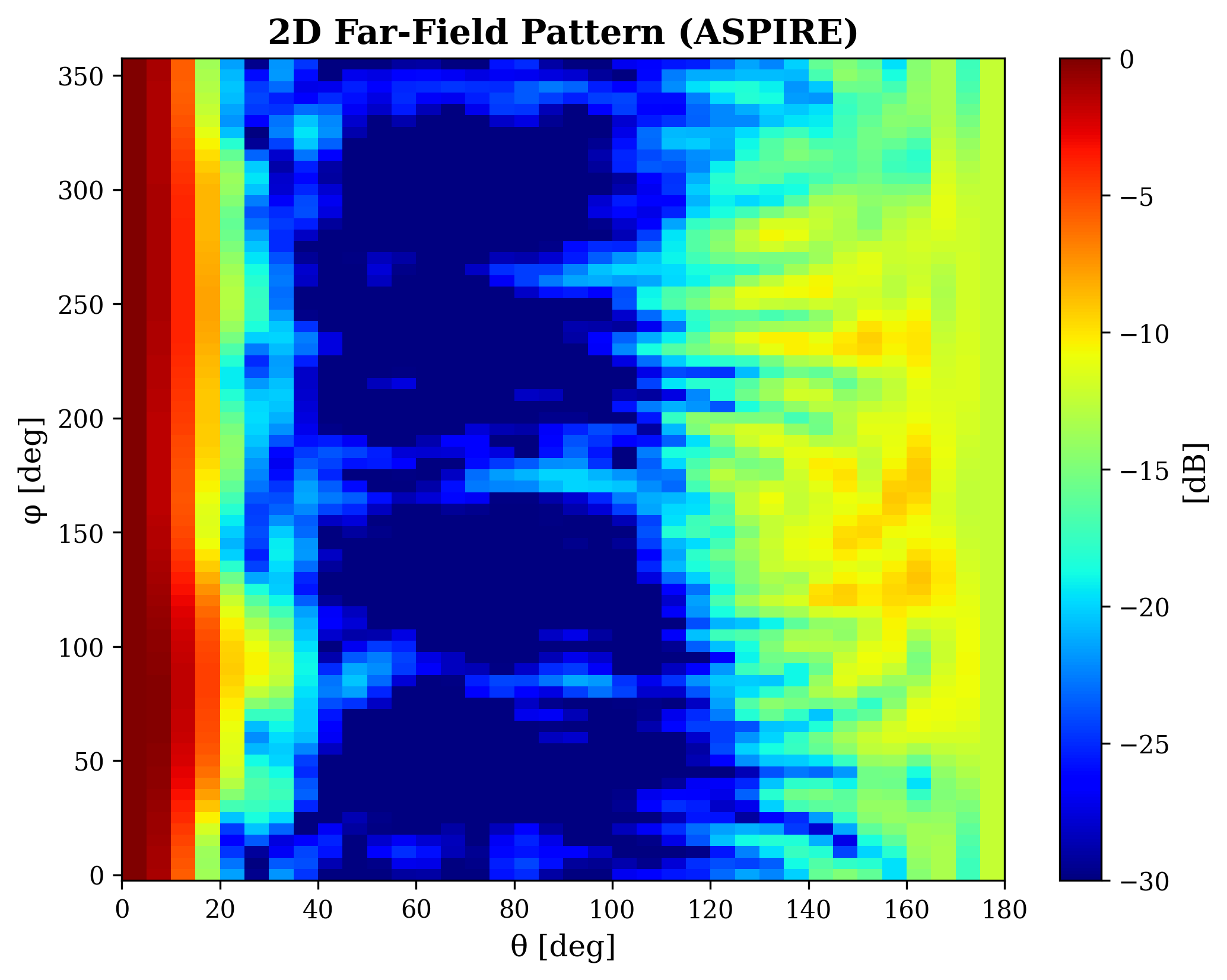}
  \caption{Two-dimensional far-field gain map ($\theta$ vs.\ $\varphi$).
    Colour scale: \SI{0}{\deci\bel} to $-30$\,dB. Dashed lines at
    $\theta = 20^{\circ}$ mark the facility-validated boundary.}
  \label{fig:2d_colormap}
\end{figure}

\label{subsec:hemisphere}%
Fig.~\ref{fig:2d_colormap} presents the reconstructed gain distribution as a
function of $\theta$ and $\varphi$ (colour scale: \SI{0}{\deci\bel} to
$-30$\,dB). Colour uniformity within the facility-validated band
($|\theta| \leq 20^{\circ}$) across all $\varphi$ confirms stable
reconstruction. The main beam is concentrated within
$\theta \in [0^{\circ}, 30^{\circ}]$, consistent with the directive AUT
character.

\subsection{Principal Plane Characterisation}%
\begin{figure}[H]
  \centering
  \includegraphics[width=\columnwidth]{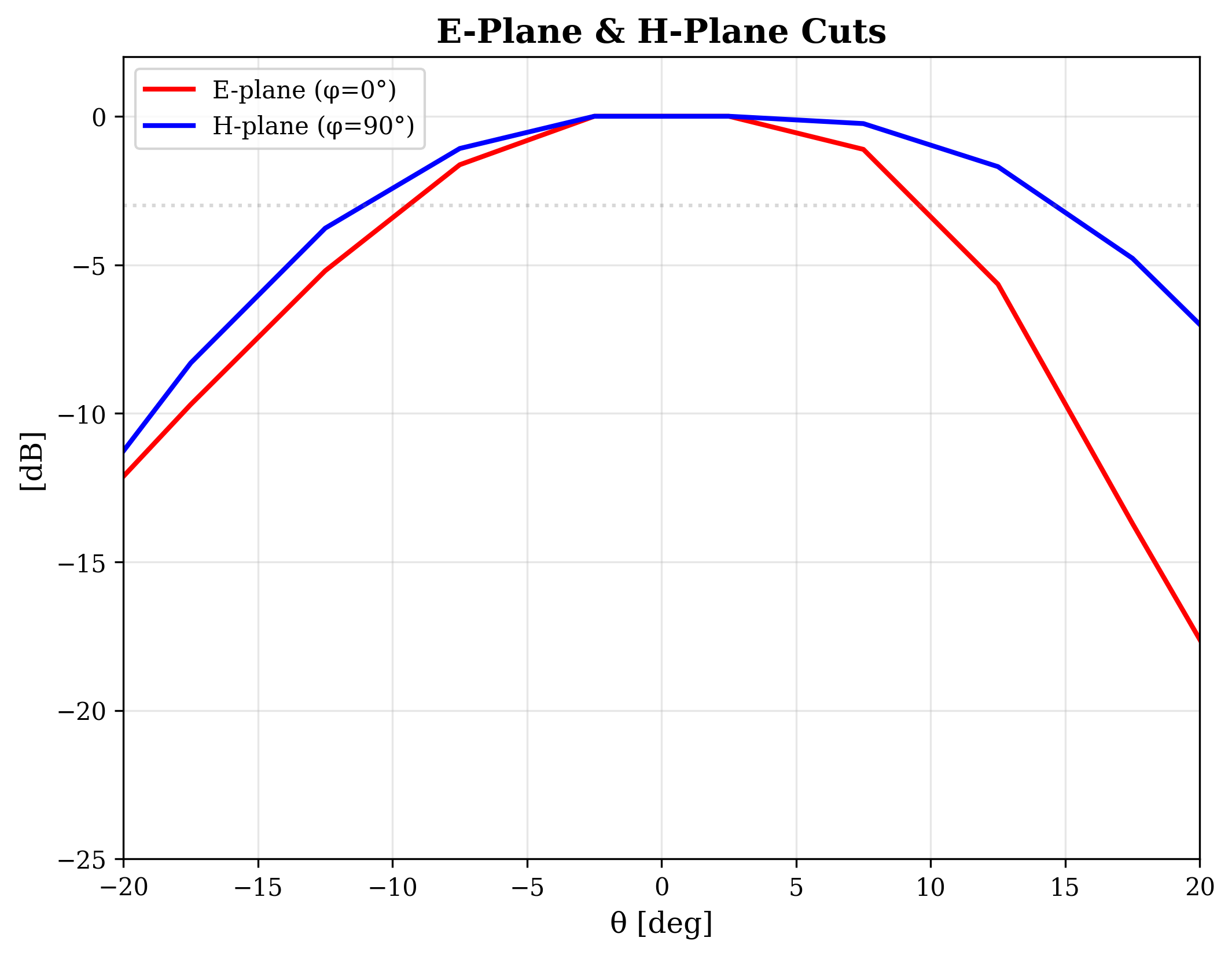}
  \caption{E-plane ($\varphi = 0^{\circ}$, solid) and H-plane
    ($\varphi = 90^{\circ}$, dashed) far-field cuts over
    $|\theta| \leq 40^{\circ}$, normalised to \SI{0}{\dBi}. The shaded band
    marks the facility-validated window $|\theta| \leq 20^{\circ}$.}
  \label{fig:eh_cuts}
\end{figure}

\label{subsec:principal_planes}%
Fig.~\ref{fig:eh_cuts} presents the E-plane ($\varphi = 0^{\circ}$) and
H-plane ($\varphi = 90^{\circ}$) cuts over $|\theta| \leq 40^{\circ}$,
derived directly from the FISTA coefficient vector.
The E-plane HPBW is $\mathrm{HPBW}_{\mathrm{E}} = 19.0^{\circ}$ and the
H-plane HPBW is $\mathrm{HPBW}_{\mathrm{H}} = 25.6^{\circ}$, with an
asymmetry of
$|\mathrm{HPBW}_{\mathrm{E}} - \mathrm{HPBW}_{\mathrm{H}}| = 6.6^{\circ}$,
consistent with the physical aperture geometry.
No facility reference is overlaid beyond $|\theta| = 20^{\circ}$; the
E-plane cut agrees with Fig.~\ref{fig:ff_comparison} to within floating-point
tolerance over the shared angular window.

\subsection{Polarisation and Cross-Polar Discrimination}%
\begin{figure}[H]
  \centering
  \includegraphics[width=\columnwidth]{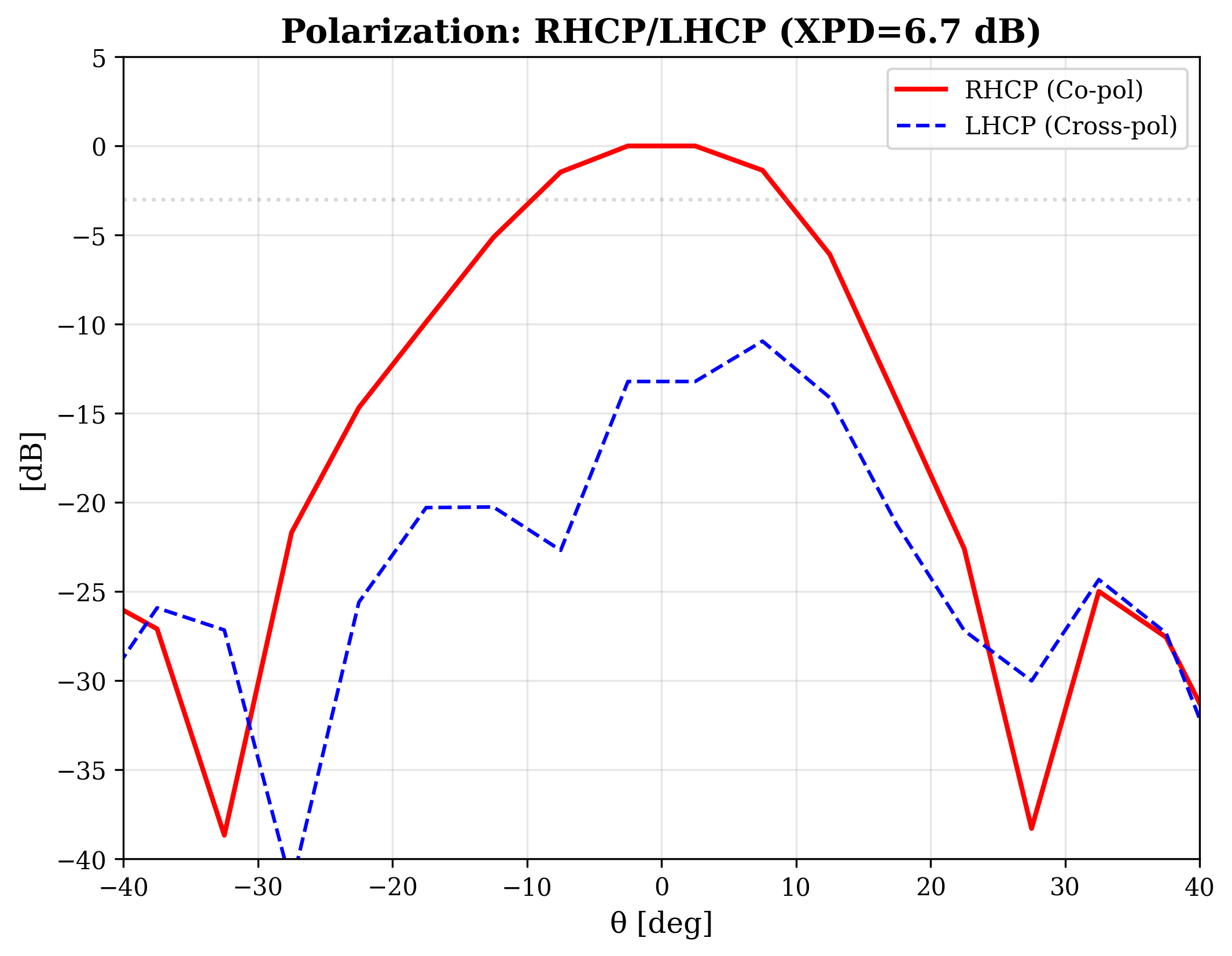}
  \caption{RHCP (solid) and LHCP (dashed) far-field patterns with XPD
    (dash-dot, right axis). Boresight XPD is \SI{6.7}{\deci\bel}
    (system-limited; UAV body depolarisation uncompensated). Dotted line:
    \SI{15}{\deci\bel} engineering-grade threshold.}
  \label{fig:polarisation}
\end{figure}

\label{subsec:polarisation}%
Fig.~\ref{fig:polarisation} presents the ASPIRE-reconstructed RHCP and LHCP
far-field patterns together with the resulting cross-polarisation discrimination
(XPD).

The boresight XPD of \SI{6.7}{\deci\bel} falls below the \SI{15}{\deci\bel}
engineering threshold, attributable to UAV body scattering rather than AUT
polarisation impurity. This result is reported as a system-level indicator
and excluded from the validated metrics in Table~\ref{tab:performance}.

\subsection{Sparse Recovery Performance and Uncertainty Quantification}%
\label{subsec:sparse_uq}%
FISTA reduced the active support from $N = 17{,}298$ to $s = 4{,}201$
coefficients (24\% sparsity, $17.3\times$ compression relative to the
rSVD rank-$k$ basis at $k = 1000$).

Hutchinson UQ yields per-coefficient standard deviations of
$\sigma \approx 10^{-3}$ for active modes and $\sigma \approx 10^{-4}$ for
suppressed modes---a one-decade separation confirming that FISTA correctly
identifies the spectral support and discards no coefficient carrying
significant posterior probability.

\subsection{Computational Performance}%
\label{subsec:compute}%
Table~\ref{tab:compute} reports measured wall-clock times for the all-dense and
hybrid MLFMM configurations on an 8\,GB unified-memory platform (Apple M-series
SoC). The hybrid configuration applies MLFMM selectively to FISTA and
Hutchinson UQ while retaining BLAS-3 \textsc{gemm} for rSVD.

\begin{table}[H]
  \caption{%
    Measured ASPIRE Wall-Clock Time: All-Dense vs.\ Hybrid MLFMM\\
    ($N\!=\!17{,}298$,\;$k\!=\!1000$,\;Apple M-series SoC,\;8\,GB unified memory)%
  }
  \label{tab:compute}
  \centering
  \renewcommand{\arraystretch}{1.25}
  \small
  {\setlength{\tabcolsep}{4pt}
  \resizebox{\columnwidth}{!}{%
  \begin{tabular}{@{}l c r r c@{}}
    \toprule
    \textbf{Stage} & \textbf{Op.} & \textbf{Dense} & \textbf{Hybrid} & \textbf{Speedup} \\
    \midrule
    rSVD ($k\!=\!1000$, 14{,}400 MVPs) & BLAS-3 & 142.3\,s & 142.3\,s & $1.00\times$ \\
    FISTA (207 MVPs)                    & MLFMM  &  69.0\,s &  59.4\,s & $1.16\times$ \\
    Hutchinson UQ ($\sim$11{,}874 MVPs) & MLFMM  & 808.7\,s & 256.6\,s & $3.15\times$ \\
    \midrule
    \textbf{Total pipeline}             & Hybrid
      & \textbf{17.0\,min}
      & \textbf{7.6\,min}
      & $\mathbf{2.23\times}$ \\
    \bottomrule
    \multicolumn{5}{@{}l}{%
      \footnotesize MVP: matrix-vector product. Hybrid = MLFMM at FISTA and UQ stages only.
    }
  \end{tabular}%
  }}
\end{table}

The hybrid pipeline achieves a $2.23\times$ end-to-end speedup
($17.0 \rightarrow 7.6$\,min). Hutchinson UQ benefits most at $3.15\times$
($808.7 \rightarrow 256.6$\,s); FISTA yields $1.16\times$ over its 207-MVP
budget. The rSVD stage retains the dense BLAS-3 kernel, as its 14{,}400
batched column-MVPs at $3.32\,\mathrm{ms\,col}^{-1}$ outperform individual
MLFMM matvecs. The full pipeline completes in under 8\,min on a laptop-class
device without server-class hardware.

\subsection{Summary of Reconstruction Performance}%
\label{subsec:summary}%
Table~\ref{tab:performance} consolidates the key quantitative reconstruction
metrics from Sections~\ref{subsec:ff_accuracy}--\ref{subsec:sparse_uq}.

\begin{table}[H]
  \renewcommand{\arraystretch}{1.3}
  \caption{ASPIRE NF-FF Reconstruction Performance at \SI{6.7125}{\GHz}}
  \label{tab:performance}
  \centering
  \begin{tabular}{@{}lccc@{}}
    \toprule
    \textbf{Metric}
      & \textbf{Facility}
      & \textbf{rSVD}
      & \textbf{rSVD\,+\,FISTA} \\
    \midrule
    HPBW, $\varphi = 0^{\circ}$
      & $18.6^{\circ}$  & $18.9^{\circ}$ & $19.0^{\circ}$ \\
    Beamwidth Error
      & ---             & $0.3^{\circ}$  & $0.4^{\circ}$  \\
    Normalised Residual
      & ---             & ---            & $1.94\%$       \\
    Active RWG Coefficients
      & ---             & $1{,}000\ (k)$ & $4{,}201\ (24\%)$ \\
    Compression Factor $\bigl(\tfrac{N}{k}\bigr)$
      & ---             & ---            & $17.3\times$   \\
    $\sigma_{\mathrm{UQ}}$, active modes
      & ---             & ---            & $\sim 10^{-3}$ \\
    $\sigma_{\mathrm{UQ}}$, suppressed modes
      & ---             & ---            & $\sim 10^{-4}$ \\
    XPD (uncalibrated)$^{\,\dag}$
      & ---             & ---            & $6.7$\,dB      \\
    \bottomrule
    \multicolumn{4}{@{}l}{%
      \footnotesize
      $^{\dag}$\,System-limited; UAV body depolarisation uncompensated.
    }
  \end{tabular}
\end{table}

\subsection{Discussion and Limitations}%
\label{subsec:discussion}%
The ASPIRE pipeline achieves a $1.94\%$ normalised residual and $0.4^{\circ}$
beamwidth error against a calibrated facility reference, confirming the viability
of sparse drone-based near-field measurement at microwave frequencies.
The rSVD\,+\,FISTA chain consistently outperforms the rSVD-only baseline,
particularly beyond the first null where sparsity enforcement suppresses
systematic sidelobe bias. The $17.3\times$ compression demonstrates that the
AUT's radiation content occupies a sparse RWG subspace.
The hybrid MLFMM strategy reduces total computation from 17.0\,min to 7.6\,min
($2.23\times$) on a laptop-class device without accuracy degradation.

Two limitations are noted: (i) ground-truth validation is bounded to
$|\theta| \leq 20^{\circ}$ by the reference facility extent, and
(ii) the XPD of \SI{6.7}{\deci\bel} falls below the \SI{15}{\deci\bel}
threshold due to uncompensated UAV body scattering. Probe polarimetric
calibration is deferred to future work (Section~\ref{sec:future}).

\section{Conclusion and Future Work}
\label{sec:future}   

This paper presented a dual-contribution drone-based antenna measurement
system: (i) a purpose-engineered hexacopter with RTK-GNSS achieving
${\leq}1$\,cm positioning accuracy and validated propulsion, endurance, and
thermal margins for field deployment at 41\,°C; and (ii) the ASPIRE
algorithm, combining RWG equivalent-current modelling, MLFMM-accelerated
forward operators, randomized SVD with Picard diagnostics, FISTA $\ell_1$
inversion with empirically selected $\lambda = 10.0$, exact debiasing, and
Hutchinson stochastic uncertainty quantification in an end-to-end pipeline.
At 6.7125\,GHz, ASPIRE recovers the far-field radiation pattern with a
residual of 1.94\% and a beamwidth error of $0.4^{\circ}$, using only 24\%
of the RWG mesh as active support.

Future work targets ASPIRE's evolution into a fully autonomous measurement
package: (i) Gaussian Process Regression with Location Errors
(GPR-LE)~\cite{mchutchon2011} for probabilistic correction of residual
RTK position uncertainty at the NF data-correction stage; (ii)
Non-Uniform Fast Fourier Transform (NUFFT)~\cite{barnett2019} spectral
synthesis for efficient far-field evaluation on irregular grids; (iii)
hybrid projection regularisation~\cite{chunggazzola2024} for adaptive
joint rank-$\lambda$ co-selection under the ill-posed compact-operator
regime; and (iv) a fully automated field-deployable package integrating
mission planning, autonomous trajectory generation, real-time data ingest,
and a calibrated uncertainty-aware far-field report as the sole user-facing
output.

\balance

\end{document}